\documentclass[aps,twocolumn,pra,showpacs,preprintnumbers,amsmath,amssymb]{revtex4}
\newcommand{\lsim} 
 {\ \raise.35ex\hbox{$<$}\kern-0.75em\lower.5ex\hbox{$\sim$}\ }
\newcommand{\gsim}
 {\ \raise.35ex\hbox{$>$}\kern-0.75em\lower.5ex\hbox{$\sim$}\ }
\newcommand{\bras}[1]{\langle#1|}
\newcommand{\kets}[1]{|#1\rangle}

\newcommand{\bra}[1]{\left<#1\right|}
\newcommand{\ket}[1]{\left|#1\right>}

\newcommand{\means}[1]{\langle#1\rangle}

\def\journal #1#2#3#4{#1 {\bf #2}, #3 (#4)}

\def\PRB{Phys.\ Rev.\ B}

\def\PRL{Phys.\ Rev.\ Lett.}

\def\JPCM{J.\ Phys.\ Cond.\ Mat.}

\def\JPSJ{J.\ Phys.\ Soc.\ Jpn.}

\usepackage{graphicx}
\usepackage{dcolumn}
\usepackage{bm}
\usepackage{amsmath}
\usepackage{times}
\usepackage[usenames]{color}

\usepackage{ulem}

\makeatletter
\def\segment#1(#2)#3({\@segment(#2)(}
\def\@segment(#1,#2)(#3,#4){%
   \@tempdima#1\p@ \advance\@tempdima#3\p@
   \divide\@tempdima\tw@
   \@tempdimb#2\p@ \advance\@tempdimb#4\p@
   \divide\@tempdimb\tw@
   \edef\@segment@temp{\noexpand\qbezier(#1,#2)%
      (\strip@pt\@tempdima,\strip@pt\@tempdimb)(#3,#4)}%
   \@segment@temp}
\makeatother

\newcommand{\QDMp}{\unitlength0.05em 
\begin{minipage}{23\unitlength}
\thicklines
 \begin{center}
 \begin{picture}(20,18)
\segment(5,0)(15,0)
\segment(20,8.66)(15,17.32)
\segment(5,17.32)(0,8.66)
 \end{picture}
 \end{center}
 \end{minipage}
}
\newcommand{\QDMm}{\unitlength0.05em 
\begin{minipage}{23\unitlength}
\thicklines
 \begin{center}
 \begin{picture}(20,18)
\segment(15,0)(20,8.66)
\segment(5,17.32)(15,17.32)
\segment(0,8.66)(5,0)
 \end{picture}
 \end{center}
 \end{minipage}
}
\newcommand{\QDMtp}{\unitlength0.05em 
\begin{minipage}{23\unitlength}
\thicklines
 \begin{center}
 \begin{picture}(20,18)
  \segment(5,0)(15,0)
  \textcolor{red}{\put(20,8.66){\circle*{6}}}
 \end{picture}
 \end{center}
 \end{minipage}
}
\newcommand{\QDMtm}{\unitlength0.05em 
\begin{minipage}{23\unitlength}
\thicklines
 \begin{center}
 \begin{picture}(20,18)
\segment(15,0)(20,8.66)
  \textcolor{red}{\put(5,0){\circle*{6}}}
 \end{picture}
 \end{center}
 \end{minipage}
}

\newcommand{\DIMER}{\unitlength0.05em 
\begin{minipage}{38\unitlength}
\linethickness{2pt}
 \begin{center}
 \begin{picture}(30,18)
  \put(5,8.66){\line(1,0){25}}
 \end{picture}
 \end{center}
 \end{minipage}
}
\newcommand{\NDIMER}{\unitlength0.05em 
\begin{minipage}{38\unitlength}
\thinlines
 \begin{center}
 \begin{picture}(30,18)
  \put(5,8.66){\line(1,0){25}}
 \end{picture}
 \end{center}
 \end{minipage}
}

\newcommand{\LSING}{\unitlength0.05em 
\begin{minipage}{23\unitlength}
\thicklines
 \begin{center}
 \begin{picture}(20,18)
  \textcolor{red}{\put(10,8.66){\circle*{6}}}
 \end{picture}
 \end{center}
 \end{minipage}
}
\newcommand{\NLSING}{\unitlength0.05em 
\begin{minipage}{23\unitlength}
\thinlines
 \begin{center}
 \begin{picture}(20,18)
  \put(10,8.66){\circle{6}}
 \end{picture}
 \end{center}
 \end{minipage}
}

\begin{document}
\title{Resonating Valence-Bond State in an Orbitally Degenerate Quantum Magnet with Dynamical Jahn-Teller Effect}
\author{Joji~Nasu} 
 \affiliation{Department of Physics, Tokyo Institute of Technology, Meguro, Tokyo 152-8551, Japan} 
\author{Sumio~Ishihara} 
 \affiliation{Department of Physics, Tohoku University, Sendai 980-8578, Japan} 
 \affiliation{JST-CREST, Sendai 980-8578, Japan}
\date{\today}
\begin{abstract}  
Short-range resonating-valence bond states in an orbitally degenerate magnet on a honeycomb lattice is studied. 
A quantum-dimer model is derived from the Hamiltonian which represents the superexchange interaction and the dynamical Jahn-Teller (JT) effect. 
We introduce two local units termed ``spin-orbital singlet dimer'', where two spins in a nearest-neighbor bond form a singlet state associated with an orbital polarization along the bond, and ``local JT singlet'', where an orbital polarization is quenched due to the dynamical JT effect. 
A derived quantum-dimer model consists of the hopping of the spin-orbital singlet dimers and the JT singlets, and the chemical potential of the JT singlets. 
We analyze the model by the mean-field approximation, and find that a characteristic phase, termed ``JT liquid phase'', where both the spin-orbital singlet dimers and the JT singlets move quantum mechanically, is realized. 
Possible scenarios for the recently observed non magnetic-ordered state in Ba$_3$CuSb$_2$O$_9$ are discussed. 
\end{abstract}


\maketitle



%
%

%




\section{Introduction}

Quantum spin liquid (QSL) state without long-range magnetic order down to low temperatures is one of the central issues in strongly correlated electron systems~\cite{Balents2010}.
A number of experimental researches to explore the QSL phenomena have been performed in a wide range of molecular-organic salts~\cite{Shimizu03,Yamashita10,Isono2014} and transition-metal compounds~\cite{Nakatsuji05,Helton07,Okamoto07} with geometrically frustrated lattices. 
Some of the materials have been analyzed theoretically based on the Heisenberg model and/or the single-band Hubbard model on frustrated lattices, which are widely accepted as the minimal models for the QSL phenomena. 

Beyond the minimal theoretical models, additional factors, which promote the QSL phenomena, have been examined for recent decades. 
An orbital degeneracy in magnetic ions is one of the candidate factors.
The orbital degree of freedom represents directions of the electronic wave function and the charge distributions. 
Some kinds of orbital alignments reduce an effective dimensionality of the magnetic interactions and disrupt the magnetic orders. 
The orbital degree of freedom also has an intrinsic frustration effect even without geometrical frustration; all bond energies in a certain orbitally ordered state cannot be minimized simultaneously. 
From these points of view, the QSL states in magnets with the orbital degeneracy, as well as the spin-orbital quantum liquid states, have been proposed theoretically~\cite{Feiner1997,Khaliullin2000,Vernay2004,Mila2007,Chen2009}, and candidate materials, such as LiNiO$_2$ and FeSc$_2$S$_4$, have been examined experimentally~\cite{Kitaoka1998,Krimmel2005}.  

The resonating valence bond (RVB) state is widely accepted as a possible ground state in low-dimensional quantum spin systems. 
This is expected to be given by a superposition of the spin singlet pairs formed in nearby sites of a crystal lattice.  
One successful theoretical treatment for the short-range RVB ground state in $S=1/2$ spin systems is known as the quantum dimer model (QDM) originally proposed in Ref.~\cite{Rokhsar1988}, in which the resonance of the valence bonds is represented by a kinetic motion of the local spin singlet dimers. 
This is extended not only to the frustrated magnets~\cite{Moesner2001} but also to a hole doped system away from the half filling, where hole kinetic energy, as well as the exchange interaction, moves the singlet dimers~\cite{Rokhsar1988}.
The QDM is also applicable to the quantum magnets with orbitally degeneracy. 
A description of the resonant spin-orbital dimers was attempted in a triangular lattice system motivated from the experiments in LiNiO$_2$~\cite{Mila2007}.
A spin-orbital system in a honeycomb lattice is another candidate, in which the spin-dimer picture gives an appropriate description for the ground state. 
Theoretical calculations based on the microscopic spin-orbital model have shown that the spin singlet dimers associated with the orbital alignment cover resonantly the honeycomb lattice~\cite{Nasu2013}.

Recently, a layered cupper oxide, Ba$_3$CuSb$_2$O$_9$, is found to be a new candidate of the QSL materials with the orbital degree of freedom~\cite{Zhou11,Nakatsuji12,Quilliam12,Ishiguro2013}.
Temperature dependence of the magnetic susceptibility shows no anomalies down to 0.2K, although the Weiss temperature is estimated to be $-55$K~\cite{Zhou11}.
In an early stage of the research, this was attributed to the $S=1/2$ spin system in a triangular lattice. 
However, detailed x-ray diffraction structure analyses revealed that Cu ions do not form a triangular lattice, but a short-range honeycomb lattice~\cite{Nakatsuji12,Katayama2014}, which requires us to search another factor to realize a QSL state. 
A promising factor is the orbital degree of freedom in a Cu$^{2+}$ ion where one hole occupies one of the degenerate $e_g$ orbitals. 
The electron spin resonance (ESR) experiments show the almost isotropic $g$ factors down to 30K~\cite{Nakatsuji12,Katayama2014}, suggesting no specific orbital alignments with static Jahn-Teller (JT) distortions, but indicate a possibility of an orbital quenching.
Some theoretical scenarios for unique roles of the orbital degree of freedom on QSL were proposed~\cite{Corboz,Nasu2013,Smerald2014}.

In this paper, we examine short-range RVB states in an orbitally degenerate magnet on a honeycomb lattice with the dynamical JT effect, motivated from the recently discovered QSL state in Ba$_3$CuSb$_2$O$_9$.  
A QDM is derived from the Hamiltonian which represents the superexchange interaction and the dynamical JT effect.
Two local objects are introduced: a spin-singlet dimer associated with the polarized orbitals along the bond, termed the spin-orbital singlet dimer, and a local JT singlet, where both the orbital polarization and the static JT distortion are quenched due to the dynamical JT effect. 
The derived QDM consists of the kinetic terms of the spin-orbital singlet dimers and the JT singlets, and the chemical potential of the JT singlets. 
The mean-field phase diagrams are obtained. 
A characteristic phase, termed the JT liquid phase, is realized due to the competition between the superexchange interactions and the dynamical JT effect; both the spin-orbital singlet dimers and the JT singlets hop on a lattice quantum mechanically. 
Relations to the experiment results in Ba$_3$CuSb$_2$O$_9$, as well as the previous theoretical works, are discussed. 

In Sec.~\ref{sec:model}, QDM is derived from the Kugel-Khomskii type superexhange interaction and the local dynamical JT effect. 
In Sec.~\ref{sec:method-result}, mean-field phase diagrams at zero temperature is calculated. Sec.~\ref{sec:discussion-summary} is devoted to the discussion and concluding remarks. 

\section{Model}
\label{sec:model}

In this section, we set up the model Hamiltonian which consists of the Kugel-Khomskii type superexchange Hamiltonian and the dynamical JT effect between the orbitals and lattice vibrations.
The QDM is derived, as an effective model of the Hamiltonian, by the perturbational expansion in terms of the superexchange interactions.

\subsection{Superexchange Interaction}\label{sec:super-inter-hamilt}
\begin{figure}[t]
\begin{center}
\includegraphics[width=\columnwidth,clip]{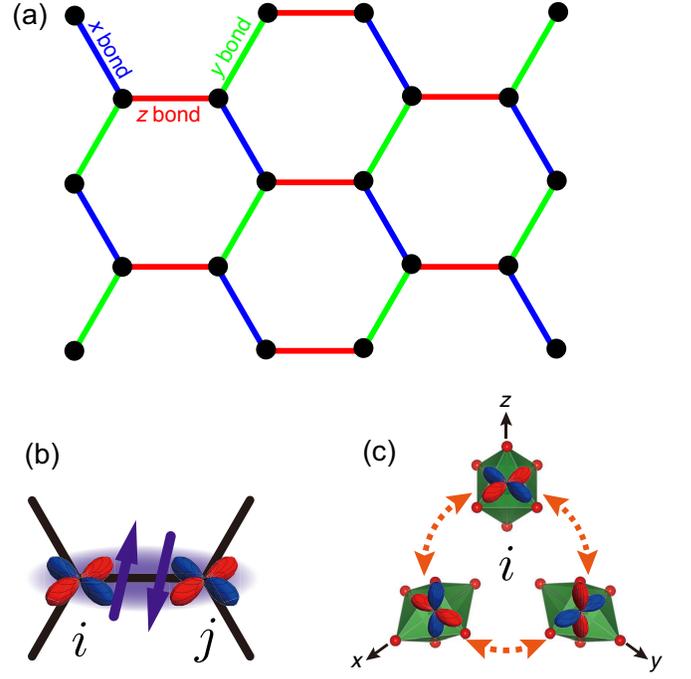}
\caption{
(a) Three kinds of NN bonds in a honeycomb lattice. 
Schematic views of (a) a spin-orbital singlet dimer $\kets{\psi^{s\tau}}_{ij;z}$, and (c) a JT singlet $\kets{\Psi_{\rm A_1}^{\rm JT}}_i$.
}
\label{fig:lattice}
\end{center}
\end{figure}

We consider the superexchange interactions between the nearest-neighbor (NN) magnetic ions on a honeycomb lattice. 
One hole occupies one of the doubly degenerate $e_g$ orbitals in each magnetic ion. 
The orbital degree of freedom is represented by the pseudo-spin operator with an amplitude of $1/2$ defined by 
\begin{align}
\bm{T}_i=\frac{1}{2}
\sum_{s\gamma\gamma'}
c_{i\gamma s}^\dagger 
\bm{\sigma}_{\gamma \gamma'} c_{i\gamma' s}, 
\end{align}
where $c_{i\gamma s}$ is an annihilation operator for a hole with orbital $\gamma$ and spin $s$. 
The superexchange interactions between the NN magnetic ions are derived from the extended-$pd$ type Hamiltonian by the perturbational calculations in terms of the electron transfer integrals. 
Detailed derivations and full expressions of the Hamiltonian were presented in Ref.~\cite{Nasu2013}.  
Here, we adopt the dominant parts of the Hamiltonian, as a minimal model for the superexchange interactions, given by 
\begin{align}
 {\cal H}^{\rm SE}&=\sum_{\means{ij}_l}{\cal H}_{ij;l}^{\rm SE}\nonumber\\
&=\sum_{\means{ij}_l}\left[J_s \bm{S}_i\cdot \bm{S}_j-J_{\tau}\tau_i^l\tau_j^l
+J_{s\tau} \bm{S}_i\cdot \bm{S}_j \tau_i^l\tau_j^l
 \right],  \label{eq:2}
\end{align}
where $\means{ij}_l$ represents the NN bond connecting sites $i$ and $j$ along a $l (=x,y,z)$ bond in a honeycomb lattice as shown in Fig.~\ref{fig:lattice}(a), and $\bm{S}_i$ is the spin operator for a hole with an amplitude of $1/2$.
We introduce the bond-dependent orbital operator defined by 
\begin{align}
 \tau_i^l=\cos \left ( \frac{2\pi n_l}{3} \right)T_i^z - \sin \left ( \frac{2\pi n_l}{3}\right )T_i^x, 
\end{align}
with $(n_z ,n_x ,n_y) = (0,1,2)$. The eigenstate of $\tau_i^l$ with the eigenvalue $+1/2$ ($-1/2$) represents the state, in which the $d_{3l^2-r^2}$ ($d_{m^2-n^2}$) orbital is occupied by a hole, where $(l, m, n)=(x, y, z)$ and their cyclic permutations. 
The wave function for the eigenvalue $-1/2$ is given by $\kets{\psi_l}=\cos(\pi n_l/3)\kets{d_{x^2-y^2}}-\sin(\pi n_l/3)\kets{d_{3z^2-r^2}}$, which represents the leaf-type orbital and is preferred due to the JT coupling, as mentioned later.
The exchange constants in Eq.~(\ref{eq:2}), $J_s$, $J_\tau$, and $J_{s \tau}$, are positive. 

The lowest-energy eigenstate of ${\cal H}_{ij;l}^{\rm SE}$ on an isolated NN bond is given by 
\begin{align}
 \kets{\psi^{s\tau}}_{ij;l}=
\frac{1}{\sqrt{2}}
\left (\kets{\uparrow\downarrow}_{ij}-\kets{\downarrow\uparrow}_{ij} \right )
\kets{\psi_l^\tau}_i\kets{\psi_l^\tau}_j,\label{eq:3}
\end{align}
where $\kets{\psi_l^\tau}_i(\equiv \kets{d_{m^2-n^2}}_i$) is the one-hole occupied state of the $d_{m^2-n^2}$ orbital at site $i$. 
A schematic view is depicted in Fig.~\ref{fig:lattice}(b).
The eigenenergy for $\kets{\psi^{s\tau}}_{ij;l}$ is $-\widetilde{\varepsilon}^{s\tau}$ with 
\begin{align}
\widetilde{\varepsilon}^{s\tau}= \left(\frac{3}{4}J_s+\frac{1}{4}J_\tau+\frac{3}{16}J_{s\tau}\right). 
\end{align}
This is a spin-singlet state and the parallel orbital alignment where the orbital and spin degrees of freedom are not entangled with each other.

\begin{figure}[t]
\begin{center}
\includegraphics[width=\columnwidth,clip]{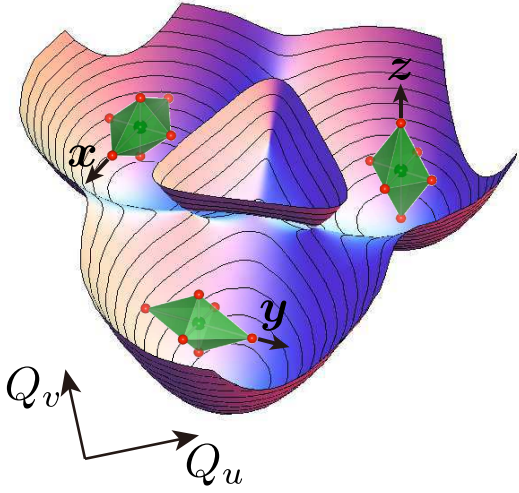}
\caption{
Adiabatic potential-energy planes of the JT coupling Hamiltonian in Eq.~(\ref{eq:1}). Lattice distortions at the three potential-energy minima in the lowest plane are also shown.
}
\label{fig:adiabatic_plane}
\end{center}
\end{figure}

\subsection{Jahn-Teller Coupling}\label{sec:jahn-tell-hamilt}

Next, we consider the dynamical JT effect between the $e_g$ orbitals and the E-symmetry vibrational modes,  $\{Q_{iu},Q_{iv}\}$, in an O$_6$ octahedron surrounding a magnetic ion at site $i$. 
We assume that the vibrations occur independently at each JT center. 
Hamiltonian is given by 
\begin{align}
{\cal H}^{\rm JT}&=\sum_i {\cal H}_i^{\rm JT} , 
\label{eq:1a}
\end{align}
with 
\begin{align}
{\cal H}^{\rm JT}_i
&=\sum_{m=(u,v)} \left (-\frac{1}{2M}\frac{\partial^2}{\partial Q_{im}^2} 
  +\frac{M\omega_0^2}{2}Q_{im}^2   \right ) \nonumber \\
&+2A(T_{i}^z Q_{iu}+T_{i}^x Q_{iv})   \nonumber \\
&+B_1(Q_{iu}^3-3Q_{iv}^2 Q_{iu}) \nonumber \\
&+B_2 \left [(Q_{iu}^2-Q_{iv}^2 )T_i^z - 2Q_{iu}Q_{iv}T_i^x \right ] \nonumber \\
&+C (Q_{iu}^2+Q_{iv}^2)^2.
\label{eq:1}
\end{align}
The first line in Eq.~(\ref{eq:1}) represents the harmonic vibrations with frequency $\omega_0$ and ionic mass $M$, and the second line represents the JT coupling with a coupling constant $A(>0)$. 
The third and succeeding lines are for the anharmonic terms where $B_1(<0)$ is the third-order anharmonic potential, $B_2(>0)$ is the quadratic JT coupling constant, and $C(>0)$ is the fourth-order anharmonic potential. 

The adiabatic energy-potential planes are calculated by neglecting the kinetic energy of the lattice vibrations. 
As shown in Fig.~\ref{fig:adiabatic_plane}, there are three potential wells on the lowest adiabatic plane, and the potential minima on the $Q_u$-$Q_v$ plane are given at the distortions, 
\begin{align}
Q_l= \cos \left( \frac{2\pi n_l}{3}\right) Q_{u} +\sin \left( \frac{2\pi n_l}{3}\right) Q_{v}, 
\end{align}
for $l=(x, y, z)$ and an amplitude $\rho_0=\sqrt{Q_{u}^2+Q_{v}^2}$. 
To describe the low-energy vibronic states on the lowest adiabatic plane, it is adequate to adopt, as a non-orthogonal basis set, the vibronic wave functions $\kets{\Psi_l^{\rm JT}}$ localized around the three potential minima at $Q_l$. 
In the crude Born-Oppenheimer approximation, the wave function is given as a product of the electronic and lattice parts as~\cite{Bersuker_text}
\begin{align}
\kets{\Psi_l^{\rm JT}}=\kets{\psi_l^\tau}\kets{\Phi_l^{\rm vib}} , 
\end{align}
where $\kets{\Phi_l^{\rm vib}}$ is the lattice wave-function localized around $Q_l$.
Although the following formulation does not depend on an explicit form of $\kets{\Phi_l^{\rm vib}}$, an example of the wave function is given by
\begin{align}
 \Phi_l^{\rm vib}={\cal N}\exp\left[-\frac{M\omega_\rho}{2}(\rho-\rho_0)^2-\frac{M\omega_\vartheta \rho_0^2}{2}\left(\vartheta-\frac{2\pi n_l}{3}\right)^2\right],
\end{align}
where $(\rho,\vartheta)$ are the polar coordinates in the $Q_u-Q_v$ plane, ${\cal N}$ is a normalization factor, and $\omega_\rho$ and $\omega_\vartheta$ are the vibrational frequencies for the $\rho$ and $\vartheta$ coordinates, respectively.
The vibronic wave-function with the A$_1$ symmetry is given by a linear combination of $\kets{\Psi_l^{\rm JT}}$ as 
\begin{align}
 \kets{\Psi_{\rm A_1}^{\rm JT}}= \frac{1}{\sqrt{3}} 
\left (\kets{\Psi_x^{\rm JT}}+\kets{\Psi_y^{\rm JT}}+\kets{\Psi_z^{\rm JT}} \right),
\end{align}
termed a JT singlet state, with the energy 
\begin{align}
 E_{\rm A_1}=\frac{\means{\Psi_{\rm A_1}^{\rm JT}|{\cal H}_i^{\rm JT}|\Psi_{\rm A_1}^{\rm JT}}}{\means{\Psi_{\rm A_1}^{\rm JT}|\Psi_{\rm A_1}^{\rm JT}}}=\frac{E_{11}+2E_{12}}{1+2S},
\end{align}
where $S=\means{\Psi_z^{\rm JT}|\Psi_x^{\rm JT}}$, $E_{11}=\means{\Psi_z^{\rm JT}|{\cal H}_i^{\rm JT}|\Psi_z^{\rm JT}}$ and $E_{12}=\means{\Psi_z^{\rm JT}|{\cal H}_i^{\rm JT}|\Psi_x^{\rm JT}}$.
A schematic view of the JT singlet is given in Fig.~\ref{fig:lattice}(c). 
In a similar way, the E-symmetry doublet wave functions are given by 
\begin{align}
  \kets{\Psi_{{\rm E}u}^{\rm JT}}&= \frac{1}{\sqrt{6}} 
\left (-\kets{\Psi_x^{\rm JT}}-\kets{\Psi_y^{\rm JT}}+2\kets{\Psi_z^{\rm JT}} \right),\\
  \kets{\Psi_{{\rm E}v}^{\rm JT}}&= \frac{1}{\sqrt{2}} 
\left (\kets{\Psi_x^{\rm JT}}-\kets{\Psi_y^{\rm JT}} \right),
\end{align}
with the energy 
\begin{align}
 E_{\rm E}=\frac{E_{11}-E_{12}}{1-S}.
\end{align}
When the quadratic JT coupling, $B_2$ term in Eq.~(\ref{eq:1}), is neglected, the doublet E states are the ground state.
On the other hand, in the case of large $B_2$, there is a possibility that the singlet A$_1$ state is the ground state~\cite{Koizumi1999}. 
We assume the singlet ground state through this paper. 
Relations to our previous work, where the doublet E states are assumed to be the ground state~\cite{Nasu2013}, will be discussed in Sec.~\ref{sec:discussion-summary}.

\subsection{Quantum Dimer Model}
\label{sec:deriv-quant-dimer}

Here, we derive QDM as an effective model of the Hamiltonian
\begin{align}
 {\cal H}={\cal H}^{\rm SE} + {\cal H}^{\rm JT} ,
\label{eq:4}
\end{align}
given in Eqs.~(\ref{eq:2}), (\ref{eq:1a}) and (\ref{eq:1}). 
We assume that the low energy states are given by configurations of the following two local units:  
i) a NN bond state along the direction $l$ given by 
\begin{align}
\kets{\phi^{\rm SO}}_{ij; l}=\kets{\psi^{s\tau}}_{ij;l}\kets{\Phi_l^{\rm vib}}_i \kets{\Phi_l^{\rm vib}}_j .  
\end{align}
This is the spin-singlet state in the $\kets{d_{m^2-n^2}}$ orbital alignment accompanied with the distortions $Q_l$ in both sites.  
This is termed a ${\it spin}$-${\it orbital}$ ${\it singlet}$ ${\it dimer}$. 
ii) a single-site state given by 
\begin{align}
\kets{\phi^{\rm JT}}_i=\kets{s}_i\kets{\Psi_{\rm A_1}^{\rm JT}} _i .
\end{align}
This is a product of the spin state ($s(=\uparrow, \downarrow)$) and the singlet vibronic state, by which an orbital polarization is quenched. 
This is termed ${\it a \ JT \ singlet}$. 
For simplicity, from now on, the spin degree of freedom at the JT singlet is neglected.
This assumption might be justified when the JT singlet sites are dilute and the magnetic interaction between them is weak. 
The spin degree of freedom in the JT singlet sites and magnetic orders will be discussed in Sec.~\ref{sec:discussion-summary}.  

\begin{figure}[t]
\begin{center}
\includegraphics[width=\columnwidth,clip]{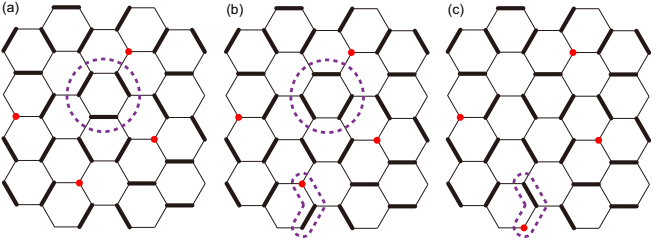}
\caption{
Three configurations of the spin-orbital singlet dimers and the JT singlets. 
Dotted circles and closed curves represent changed portions between (a) and (b), and (b) and (c), respectively.  
}
\label{fig:resonant}
\end{center}
\end{figure}
As examples, three configurations of the spin-orbital singlet dimers and the JT singlets are shown in Fig.~\ref{fig:resonant}, where $\kets{\phi^{\rm SO}}_{ij;l}$ and $\kets{\phi^{\rm JT}}_i$ are represented by symbols $\kets{\DIMER}_{ij}$ and $\kets{\LSING}_i$, respectively. 
First, we derive the matrix elements of the superexchange interactions.
A bond energy for the spin-orbital singlet dimer is given by 
\begin{align}
 {\cal H}_{ij;l}^{s\tau}\kets{\DIMER}=-\widetilde{\varepsilon}^{s\tau}\kets{\DIMER} .
\end{align}
A bond, in which the $i$ and $j$ sites are involved in different spin-orbital singlet bonds, is represented by $\kets{\NDIMER}$, and its bond energy is $\means{\NDIMER|{\cal \widetilde H}^{s\tau}_{ij; l}|\NDIMER}=-J_\tau/16\means{\NDIMER|\NDIMER}$. 
On the other hand, at least either one of $i$ and $j$ sites is occupied by the JT singlet state, the superexchange interaction energy is zero. 
By modifying the superexchange interaction as 
\begin{align}
 \widetilde{\cal H}_{ij;l}^{s\tau}={\cal H}_{ij;l}^{s\tau}+\frac{J_\tau}{16}(1-\hat{n}_i^{\rm JT})(1-\hat{n}_j^{\rm JT}),
\end{align}
where $\hat{n}_i^{\rm JT}$ is the number operator of the JT singlet at site $i$, the matrix elements are given by   
\begin{align}
\means{\DIMER|{\cal \widetilde H}_{ij;l}^{s\tau}|\DIMER}
=-\varepsilon^{s\tau}\means{\DIMER|\DIMER},\\
\means{\DIMER|{\cal \widetilde H}_{ij;l}^{s\tau}|\NDIMER}
=-\varepsilon^{s\tau} \means{\DIMER|\NDIMER} ,  
\end{align}
and others are zero. 
We define $\varepsilon^{s\tau}=\widetilde{\varepsilon}^{s\tau}-\frac{1}{16}J_\tau=\frac{3}{4}J_s+\frac{3}{16}J_\tau+\frac{3}{16}J_{s\tau}$. 

The matrix elements for the JT coupling term are calculated in the similar way. 
A site, which belongs to the spin-orbital singlet dimer, is represented by $\kets{\NLSING}$, 
and the effective JT coupling Hamiltonian is modified as 
\begin{align}
 \widetilde{\cal H}_i^{\rm JT}={\cal H}_i^{\rm JT}-E_{11} . 
\end{align} 
Then, the matrix elements are given by 
\begin{align}
 \bras{\LSING}\widetilde{\cal H}_{i}^{\rm JT}\kets{\LSING}&=-\varepsilon^{\rm JT}\means{\LSING |\LSING},\end{align}
\begin{align}
 \bras{\LSING}\widetilde{\cal H}_{i}^{\rm JT}\kets{\NLSING}&=-\varepsilon^{\rm JT}\means{\LSING |\NLSING} , 
 \end{align}
and others are zero. 
We define $\varepsilon^{\rm JT}=-E_{{\rm A}_1}+E_{11}$. 

By using the above formula, we rewrite the Hamiltonian as 
\begin{align}
 {\cal H}=\widetilde{\cal H}^{s\tau}+\widetilde{\cal H}^{\rm JT}+\frac{3J_\tau}{16}\hat{N}_{\rm JT}-\frac{J_\tau}{16}\sum_{<ij>}\hat{n}_i^{\rm JT}\hat{n}_j^{\rm JT},\label{eq:5}
\end{align}
with the total number of the JT singlet given by  
\begin{align}
 \hat{N}_{\rm JT}=\sum_i \hat{n}_i^{\rm JT}.
\end{align}
This Hamiltonian is divided into the diagonal part 
\begin{align}
 {\cal H}_0&=-\mu\hat{N}_{\rm JT}-U\sum_{<ij>}\hat{n}_i^{\rm JT}\hat{n}_j^{\rm JT}, 
\end{align}
and the off-diagonal part 
\begin{align}
{\cal H}_1&={\cal H}-{\cal H}_0,
\end{align}
where  we define the chemical potential, $\mu=-\varepsilon^{s\tau}/2+\varepsilon^{\rm JT}-3J_\tau/16$, and the inter-site interaction, $U=J_\tau/16$, for the JT singlets.

Derivation of the QDM is followed by the procedure proposed in Refs.~\cite{Rokhsar1988,Ralko2009}. 
From the non-orthogonal set $\ket{\Psi_C}$ with a configuration $C$ of the spin-orbital singlet dimers and the JT singlets, 
the orthonormal basis set is obtained as $\ket{\Phi_C}=\sum_{C'} ({\cal O}^{-1/2})_{C'C} \kets{\Psi_{C'}}$
where ${\cal O}_{CC'}=\means{\Psi_C|\Psi_{C'}}$. 
Then, the effective Hamiltonian is derived from the original Hamiltonian ${\cal H}$ is given by  
\begin{align}
\left ({\cal H}_{{\rm eff}} \right)_{AB}=\sum_{A' B'} ({\cal O}^{-1/2})_{AA'} {\cal H}_{A'B'} ({\cal O}^{-1/2})_{B'B} .
\end{align}
To obtain the expression of ${\cal H}_{\rm eff}$, the overlap matrix, ${\cal O}$, between two configurations are required to be calculated. 
Let us consider, first, the two configurations $\kets{\Psi_1}$ and $\kets{\Psi_2}$ shown in Fig.~\ref{fig:resonant}(a) and (b), respectively, where changed portions marked by broken circles are represented by the states $\kets{\QDMp}$ and $\kets{\QDMm}$. 
The two configurations are connected by hoppings of the three spin-orbital singlet dimers. 
An overlap matrix element between the two is given by  
$\means{\QDMp|\QDMm} \equiv {\cal O}_{\Psi_1 \Psi_2}=\alpha^4 S^6$ where $\alpha=1/\sqrt{2}$.
Another configuration, termed $\kets{\Psi_3}$, is shown in Fig.~\ref{fig:resonant}(c), 
where the number of sites involved in a changed portion from $\kets{\Psi_1}$ is more than six. 
An overlap matrix element between $\kets{\Psi_1}$ and $\kets{\Psi_3}$ is of the higher order of $\alpha$ than $\alpha^4$. 
Then, ${\cal O}$ is expanded by the parameter $\alpha$ as 
${\cal O}=1+\alpha^4S^6\hat{\omega}+\cdots$ 
with  
\begin{align}
 \hat{\omega}=\sum_{\rm conf.}\ket{\QDMp}\bra{\QDMm}+{\rm H.c.},
\end{align}
where a symbol $\sum_{\rm conf.}$ implies that a summation is taken for all hexagons in a honeycomb lattice. 
In the similar way, the off-diagonal term of the Hamiltonian is expanded as
${\cal H}_1=-3\varepsilon^{s\tau}\alpha^4S^6\hat{\omega}+\cdots$.
By using the expressions of ${\cal O}$ and ${\cal H}_1$, the effective Hamiltonian for the spin-orbital singlet dimer is given by 
\begin{align}
{\cal O}^{1/2}
{\cal H}
{\cal O}^{1/2}
={\cal H}_0-3 \varepsilon^{s\tau}\alpha^4 S^6 \hat{\omega}+3 \varepsilon^{s \tau} \alpha^8 S^{12} \hat{\omega}^2  
\cdots . 
\end{align}
The configurations $\kets{\Psi_2}$ and $\kets{\Psi_3}$ shown in Figs.~\ref{fig:resonant} (b) and (c), respectively, are connected due to a hopping of the JT singlet. 
Changed portions between the two configurations marked by dotted curves are represented by the local states 
$\ket{\QDMtp}$ and $\ket{\QDMtm}$. 
Then, the off-diagonal matrix element between the two is obtained as 
$\bras{\QDMtp} {\cal H}_1 \ket{\QDMtm}=
-(\varepsilon^{s\tau}+\varepsilon^{\rm JT})
\means {\QDMtp | \QDMtm}
$
with 
$\means {\QDMtp | \QDMtm}=\frac{2}{3}\alpha^2(1+2S)^2 S$.

Up to the order of $\alpha^8$, we obtain QDM as an effective model of the spin-orbital coupled system with the dynamical JT effect: 
\begin{align}
 {\cal H}_{\rm QDM}=&-t\sum_{\rm conf.}\left(\ket{\QDMp}\bra{\QDMm}+{\rm H.c.}\right)
\nonumber \\
&+V\sum_{\rm conf.}\left(\ket{\QDMp}\bra{\QDMp}+\ket{\QDMm}\bra{\QDMm}\right)\nonumber\\
&-U\sum_{<ij>}\hat{n}_i^{\rm JT}\hat{n}_j^{\rm JT}-\mu\hat{N}_{\rm JT} \nonumber \\
&-t'\sum_{\rm conf.'}\left(\ket{\QDMtp}\bra{\QDMtm}+{\rm H.c.}\right) ,\label{eq:7}
\end{align}
where $t=3\varepsilon^{s\tau}\alpha^4S^6$, $V=3\varepsilon^{s\tau}\alpha^8S^{12}$, and $t'=\frac{2}{3}\alpha^2(1+2S)^2 S$. 
A symbol $\sum_{\rm conf.'}$ represents that a summation is taken for the NN three sites in a honeycomb lattice. 
From now on, to avoid complexity due to a number of interactions in the Hamiltonian, we focus on the representative three terms, the $t$-term, $\mu$-term and $t'$-term, representing the superexchange interaction, the JT coupling, and the dynamical JT effect, respectively. 
Finally, we have a simple version of the QDM given by 
\begin{align}
 {\cal H}_{\rm QDM}=
&-t\sum_{\rm conf.}\left(\ket{\QDMp}\bra{\QDMm}+{\rm H.c.}\right)\nonumber\\
&-t'\sum_{\rm conf.'}\left(\ket{\QDMtp}\bra{\QDMtm}+{\rm H.c.}\right)-\mu\hat{N}_{\rm JT}.\label{eq:8}
\end{align}
The first and second terms represent the kinetic energies of the spin-orbital singlet dimers and the JT singlets, respectively, 
and the third term is for the chemical potential of the JT singlets.  
This model is similar to the QDM proposed in a hole-doped antiferromagnetic Mott insulator~\cite{Rokhsar1988}. 

\section{Phase Diagram}\label{sec:method-result}

\begin{figure}[t]
\begin{center}
\includegraphics[width=\columnwidth,clip]{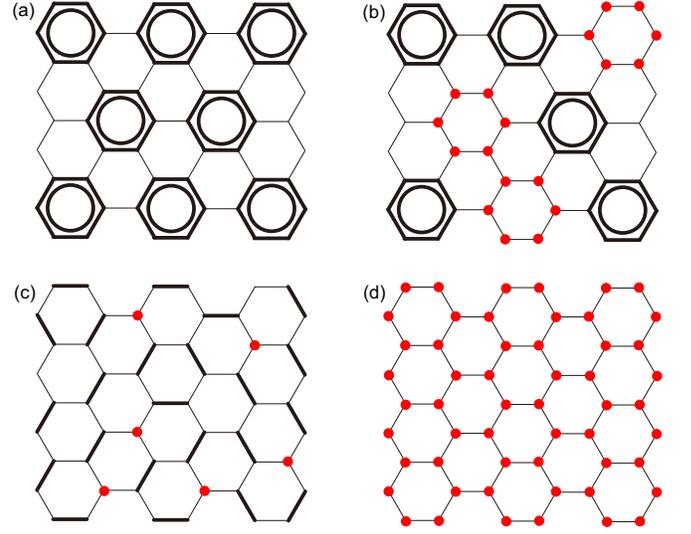}
\caption{
Four mean-field states. (a) ``Plaquette state'', where the benzene-like resonating states on hexagons are aligned on a honeycomb lattice. (b) ``Diluted plaquette state'', where some resonating states in (a) are replaced by the JT singlets. (c) ``JT liquid state'', where both the spin-orbital singlet dimer and the JT singlets hop. 
(d) ``JT singlet state'', where all sites are occupied by the JT singlets. 
}
\label{fig:mf}
\end{center}
\end{figure}
We analyze the QDM in Eq.~(\ref{eq:8}) by using the mean-field approximation, and obtain the phase diagrams at zero temperature. 
Four mean-field states shown in Fig.~\ref{fig:mf} are considered.  
These are characterized by the number density of the JT singlet, $n_{\rm JT}=\means{N_{\rm JT}}/N$, and $t'$. 
A phase shown in Fig.~\ref{fig:mf}(a) is termed a ``plaquette state'', where $n_{\rm JT}=0$ and 
hexagons occupied by the spin-orbital singlet dimers resonantly are aligned in a honeycomb lattice. 
This state is stabilized by the kinetic energy of the spin-orbital singlet dimer, i.e. the first term in Eq.~(\ref{eq:8})~\cite{Moesner2001a}.
On the other side at $n_{\rm JT}=1$, all sites are occupied by the JT singlets [see Fig.~\ref{fig:mf}(d)]. 
In between, two candidate states are shown in Figs.~\ref{fig:mf}(b) and (c), which are termed ``diluted plaquette state'' and ``JT liquid state'', respectively. 
In the diluted plaquette state, some hexagons occupied by the spin-orbital singlet-dimers in the plaquette state are replaced by the JT singlets, and all local units are localized. 
On the other hand, in the JT liquid state, the spin-orbital singlet dimers hop due to the kinetic energy of the JT singlets, i.e. the second term of Eq.~(\ref{eq:8}).  A configuration shown in Fig.~\ref{fig:mf}(c) is a snapshot.
\begin{figure}[t]
\begin{center}
\includegraphics[width=\columnwidth,clip]{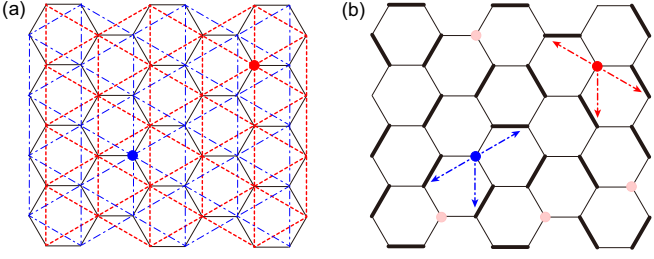}
\caption{
(a) Two sublattices (dotted and dashed-dotted lines) in a honeycomb lattice. 
(b) Possible hopping processes of  the JT singlets by the $t'$-term in Eq.~(\ref{eq:8}). 
Hopping directions depend on configurations of the spin-orbital singlet dimers surrounding the JT singlets.
}
\label{fig:sublattice}
\end{center}
\end{figure}

Energy of each phase is calculated in the QDM Hamiltonian. 
In the diluted-plaquette state, where the $t'$-term is irrelevant, the energy is given by  
\begin{align}
 \frac{E_{\rm DP}}{N}=-\frac{t}{6}(1-n_{\rm JT})-\mu n_{\rm JT}.\label{eq:10}
\end{align}
In the JT liquid state,  the $t$-term in Eq.~(\ref{eq:8}) is irrelevant, and the kinetic-energy gain of the JT singlets stabilizes the state. 
As shown in Fig.~\ref{fig:sublattice}(a), the canonical two sublattices given by connecting the next NN sites in a honeycomb lattice are termed $A$ and $B$, and the JT singlets hop to one of the three NN sites inside of the same sublattice. 
For simplicity, we assume that a JT singlet hops to one of the three directions, and neglect a fact that the hopping direction is ruled by the configurations of the spin-orbital singlet dimers around it (see Fig.~\ref{fig:sublattice}(b)). 
Then, the kinetics of the JT singlets is mapped onto the hard-core boson model, where two bosons hop  independently in each sublattice, given by 
\begin{align}
 {\cal H}_{\rm JL}=&-t'\sum_{<ij>\in A} (a_i^\dagger a_j + {\rm H.c.})-\mu\sum_{i\in A}a_i^\dagger a_i\nonumber\\
&-t'\sum_{<ij>\in B} (b_i^\dagger b_j + {\rm H.c.})-\mu\sum_{i\in B}b_i^\dagger b_i,
\end{align}
where $a_i$ and $b_i$ are the boson operators defined in sublattices $A$ and $B$, respectively. 
Symbols $\sum_{<ij>\in A(B)}$ imply summations for three of the six NN bonds in the triangular lattices. 
The total number of the JT singlets is equal to  that of the bosons: 
\begin{align}
\hat{N}_{\rm JT}=\sum_{i\in A}a_i^\dagger a_i+\sum_{i\in B}b_i^\dagger b_i. 
\end{align}
It is well known that the hard-core bosons are mapped onto the $S=1/2$ spin systems by using the relations, 
\begin{align}
 a_i^\dagger=S^{A+}_i,\  a_i=S^{A-}_i,\ a_i^\dagger a_i= \frac{1}{2}+S_i^{Az} , \\
 b_i^\dagger=S^{B+}_i,\ b_i=S^{B-}_i,\ b_i^\dagger b_i= \frac{1}{2}+S_i^{Bz}.
\end{align}
Then, the effective Hamiltonian for the JT liquid state is rewritten as 
\begin{align}
 {\cal H}_{\rm JL}=&-2t'\sum_{<ij>\in A} (S_i^{Ax} S_j^{Ax}+S_i^{Ay} S_j^{Ay})-\mu\sum_{i\in A}\left(\frac{1}{2}+S_i^{Az}\right) 
\nonumber\\
&-2t'\sum_{<ij>\in B} (S_i^{Bx} S_j^{Bx}+S_i^{By} S_j^{By})-\mu\sum_{i\in B}\left(\frac{1}{2}+S_i^{Bz}\right) . 
\end{align}
This is the two independent XY models on the triangular lattices with the transverse fields along $z$. 
We note that the interaction is antiferromagnetic, i.e. $t'[=\frac{2}{3}\alpha^2 (1+2S)^2 S] < 0$, because $S<0$.  
\begin{figure}
\begin{center}
\includegraphics[width=0.9\columnwidth,clip]{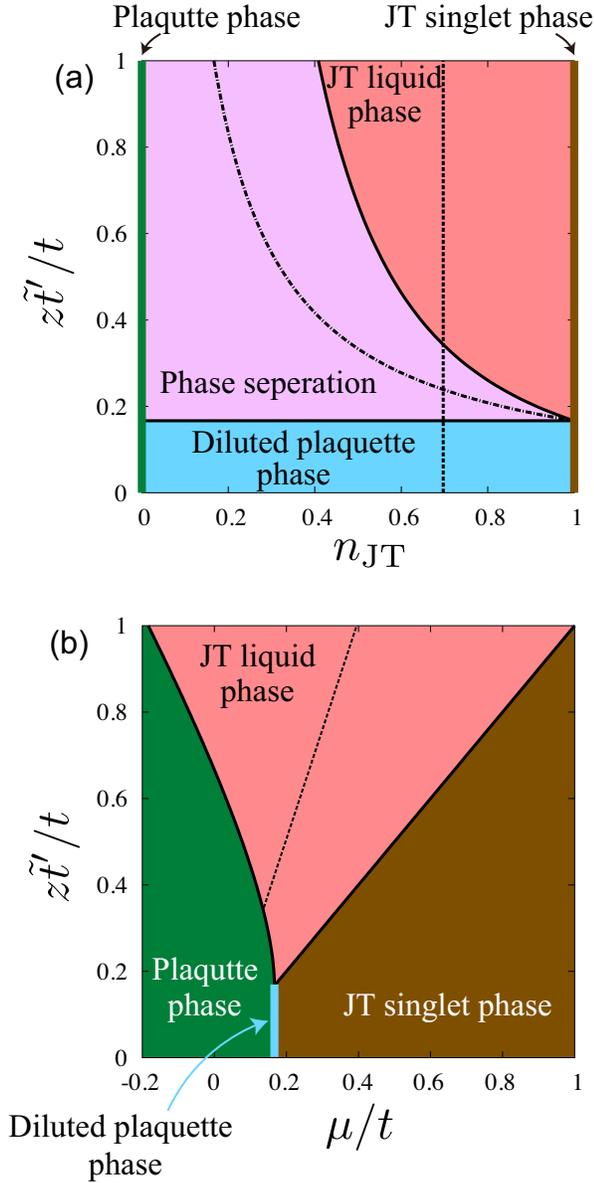}
\caption{
(a) Mean-field phase diagrams on the $n_{\rm JT}$-$\widetilde{t}'$, and (b) $\mu$-$\widetilde{t}'$ planes. 
A dashed-dotted line in (a) represents the phase boundary between the diluted plaquette and JT liquid phase, when the phase separation is not taken into account. 
Dashed lines represent the percolation threshold for a honeycomb lattice, $n_{\rm JT}=0.969$.
}
\label{fig:phase}
\end{center}
\end{figure}

As a plausible mean-field ground state of ${\cal H}_{\rm JL}$, we assume a cone spin structure, in which the cone angle is $2\theta$ and the 120-degree structure is in the $S^x-S^y$ plane.  
Energy is given by
\begin{align}
 \frac{E_{\rm JL}}{N}&=-\frac{z\widetilde{t}'}{4}\sin^2\theta-\frac{\mu}{2}(1+\cos\theta) \nonumber \\
&=-z\widetilde{t}' n_{\rm JT} (1-n_{\rm JT})-\mu n_{\rm JT},\label{eq:9}
\end{align}
where $\widetilde{t}'=|t'|/2$, $z=3$ which is an effective coordination number, and a relation $n_{\rm JT}=(1+\cos \theta)/2$ is used.

\begin{figure}[t]
\begin{center}
\includegraphics[width=0.9\columnwidth,clip]{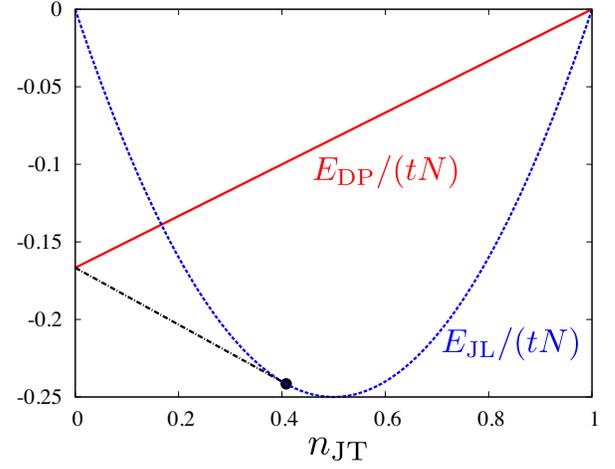}
\caption{
Energies of the diluted plaquette (a red line) and JT liquid (a blue line) states as functions of the number density of the JT singlets. 
Dashed-dotted line represents the energy given by the Maxwell construction method.
A parameter value is chosen to be $z{\widetilde t}'/t=1$.
}
\label{fig:maxwell}
\end{center}
\end{figure}
The phase diagram on the $n_{\rm JT}-t'$ plane is calculated by using Eqs.~(\ref{eq:10}) and~(\ref{eq:9}), and is presented in Fig.~\ref{fig:phase}(a). 
At $n_{\rm JT}=0$ and $1$, the plaquette phase and the JT singlet phase appear,  respectively. 
In between, the diluted plaquette phase and the JT liquid phase compete with each other. 
Energies of the two phases are compared in Fig.~\ref{fig:maxwell} where $z{\widetilde t}'/t$ is fixed to be 1. 
Two energies cross with each other at $n_{\rm JT}/(z\widetilde{t}'/t)=1/6$. 
It is shown that a phase separation between the two appears between $0< n_{\rm JT}<\sqrt{t/(6z\widetilde{t}')}$. 
By taking the phase separation into account, (see bold lines in Fig.~\ref{fig:phase}(a)), the diluted plaquette phase is restricted in a region of $z\widetilde{t}'/t<1/6$, and a phase separation governs a large parameter space. 
The phase diagram plotted on a $\mu-\widetilde{t}'$ plane is shown in Fig.~\ref{fig:phase}(b). 
The dilute plaquette phase only appears at $\mu/t=1/6$ and $z{\widetilde t}'/t<1/6$. 

Finally, we recall the relations for the parameters $t=3\varepsilon^{s\tau}\alpha^4S^6$, $t'=\frac{2}{3}\alpha^2 (1+2S)^2 S$, and $\mu=\varepsilon^{\rm JT}-\frac{1}{2}\varepsilon^{s\tau}-\frac{3}{16}J_\tau$, where $\varepsilon^{s\tau}[=\frac{3}{4}J_s+\frac{3}{16}J_\tau+\frac{3}{16}J_{s\tau}]$ is the spin-orbital singlet bond energy, and $S[=\means{\Psi^{\rm JT}_z|\Psi^{\rm JT}_x}]$ is the overlap integral of the vibrational wave function in different potential minima. 
From the viewpoint of the original model Hamiltonian in Eqs.~(\ref{eq:4}), results in Fig.~\ref{fig:phase} are interpreted that the large energy gains of the JT effect and the superexchange interaction realize the plaquette and JT singlet phases, respectively, and in between the two phases, the diluted plaquette and JT liquid phases are stabilized due to small and large contributions from the dynamical JT effects, respectively. 

\section{Discussion and Summary}
\label{sec:discussion-summary}

First, we discuss relations of the present QDM to our previous work in the spin-orbital-lattice coupled system~\cite{Nasu2013}. 
As explained in Sec.~\ref{sec:jahn-tell-hamilt}, there are two possible ground states in the local JT Hamiltonian, ${\cal H}^{\rm JT}$, introduced in Eq.~(\ref{eq:1}); the doublet E and singlet A$_{1}$ states. 
Relative stability of the two states is determined by the quadratic JT coupling, $B_2$, in Eq.~(\ref{eq:1}); the E (A$_{1}$) states are the ground state in the case of small (large) $B_2$.
In the previous study, we assumed that the doublet E states are the ground state. 
It was proposed that the spin-orbital resonant state, where the spin-orbital singlet dimers hop on a honeycomb lattice without the transnational symmetry breaking, is realized by the dynamical JT effect, and was considered as a candidate state of QSL.  
In the present work, we assume another case that the A$_{1}$ state is the ground state.
This state is represented in the QDM by a mobile local singlet, which is similar to a mobile hole carrier in the QDM for a hole doped antiferromagnetic system~\cite{Rokhsar1988}.
By the mean-field calculation, we show that the JT liquid state is stabilized by the competition between the superexchange interaction and the dynamical JT effect. 
There is another difference from the previous study; only the dominant terms of the superexchange interactions are taken into account in Eq.~(\ref{eq:2}), instead of all terms of the superexchange interactions derived in Ref.~\cite{Nasu2013}.  
The remaining terms bring about the interactions between the spin-orbital singlet dimers~\cite{Nasu2013}, and may replace the plaquette phase by the valence bond solid state for the spin-orbital singlet dimers in a region of small $z\widetilde{t}'/t$ in Fig.~\ref{fig:phase}.

Second, we discuss relations to the experimental results in Ba$_3$CuSb$_2$O$_9$. 
The two states mentioned above, i.e., the spin-orbital resonant state proposed in Ref.~\cite{Nasu2013} and the JT liquid state shown in Fig.~\ref{fig:mf}(c), are the two possible scenarios for the observed no magnetic-ordered state where the orbital polarization might be also quenched. 
Let us focus on the observed temperature dependence of the magnetic susceptibility, decomposed into the gapped and paramagnetic Curie components. 
In the scenario of the spin-orbital resonant state, the gapped and paramagnetic components are attributed to the spin-orbital singlet dimers and the so-called orphan spins, respectively.
On the other hand, in the JT liquid-state scenario, the gapped component is owing to the spin-orbital singlet dimers. 
When the number density of the JT singlets is less than the percolation threshold, i.e., $n_{{\rm JT}c}=0.696$ on a honeycomb lattice~\cite{Djordjevic1982}, it is reasonable to assume that the spin degree of freedom on the JT singlets, which is not taken into account explicitly so far, is responsible for a paramagnetic component. 
The broken lines in Figs.~\ref{fig:phase}(a) and (b) represent $n_{{\rm JT}}=n_{{\rm JT}c}$, below which two component magnetic excitations are explained in the JT liquid state. 
Weak anisotropy in the $g$ factor observed in the ESR experiments is also compatible to both the two scenarios. 
In the spin-orbital resonant state, the spin-orbital singlet dimers hop quantum mechanically~\cite{Nasu2013}, and averaged populations in the three equivalent orbitals are equal with each other in the time scale slower than the hopping energy of the spin-orbital singlet dimers. 
In the JT liquid-state scenario, the orbital polarization is also quenched due to the motion of both the spin-orbital singlet dimers and the JT singlets. 

In conclusion, we derive the QDM for a $S=1/2$ quantum spin system associated with the $e_g$ orbital degree of freedom and the dynamical JT effect.
In order to construct the QDM, two local units are introduced; the spin-orbital singlet dimer, where the two spins in a NN bond form a singlet state associated with the orbital polarization along the bond, and the local JT singlet, where the orbital polarization is quenched due to the dynamical JT effect. 
The QDM consists of the hoppings of the spin-orbital singlet dimer and the JT singlet, and the chemical potential of the JT singlet. 
Mean-field calculations reveal that the JT liquid phase, where both the spin-orbital singlet dimers and the local JT singlets hop quantum mechanically, is realized by a competition between the superexchange interaction and the dynamical JT effect.
We propose that this phase is a candidate state for the non magnetic-ordered phase observed in Ba$_3$CuSb$_2$O$_9$. 
Present study provides a new theoretical framework for the short-range RVB state in orbitally degenerated quantum magnets. 

\begin{acknowledgments}
We thank K.~Penc, S. Nakatsuji, H. Sawa, M. Hagiwara, and Y. Wakabayashi for helpful discussions. 
This work was supported by JSPS KAKENHI Grant Numbers 26287070. 
Some of the numerical calculations were performed using the supercomputing facilities at ISSP, the University of Tokyo.
JN is supported by the global COE program ``Weaving Science Web beyond Particle-Matter Hierarchy'' and the Japan Society for the Promotion of Science.
\end{acknowledgments}


\begin{thebibliography}{99} 





\bibitem{Balents2010}
L.~Balents,
\journal{Nature}{464}{199}{2010}.

\bibitem{Shimizu03}
Y.~Shimizu, K.~Miyagawa, K.~Kanoda, M.~Maesato, and G.~Saito,
\journal{\PRL}{91}{107001}{2003}.

\bibitem{Yamashita10}
M.~Yamashita, N.~Nakata, Y.~Senshu, M.~Nagata, H.~M.~Yamamoto, R.~Kato, T.~Shibauchi, and Y.~Matsuda,
\journal{Science}{328}{1246}{2010}.

\bibitem{Isono2014}
T. Isono, H. Kamo, A. Ueda, K. Takahashi, M. Kimata, H. Tajima, S. Tsuchiya, T. Terashima, S. Uji, and H. Mori,
\journal{\PRL}{112}{177201}{2014}.



\bibitem{Nakatsuji05}
S.~Nakatsuji, Y.~Nambu, H.~Tonomura, O.~Sakai, S.~Jonas, C.~Broholm, H.~Tsunetsugu, Y.~Qiu, and Y.~Maeno,
\journal{Science}{309}{1697}{2005}.

\bibitem{Helton07}
J.~S.~Helton, K.~Matan, M.~P.~Shores, E.~A.~Nytko, B.~M.~Bartlett, Y.~Yoshida, Y.~Takano, A.~Suslov, Y.~Qiu, J.-H.~Chung, D.~G.~Nocera, and Y.~S.~Lee,
\journal{\PRL}{98}{107204}{2007}.

\bibitem{Okamoto07}
Y.~Okamoto, M.~Nohara, H.~Aruga-Katori, and H.~Takagi,
\journal{\PRL}{99}{137207}{2007}.



\bibitem{Feiner1997}
L. F. Feiner, A. M. Ole\'s, and J. Zaanen,
\journal{\PRL}{78}{2799}{1997}.

\bibitem{Khaliullin2000}
G. Khaliullin and S. Maekawa,
\journal{\PRL}{85}{3950}{2000}.



\bibitem{Vernay2004}
F. Vernay, K. Penc, P. Fazekas, and F. Mila,
\journal{\PRB}{70}{014428}{2004}.


\bibitem{Mila2007}
 F. Mila, F. Vernay, A. Ralko, F. Becca, P. Fazekas, and K. Penc,
\journal{\JPCM}{19}{145201}{2007}.



\bibitem{Chen2009}
G. Chen, L. Balents, and A. P. Schnyder,
\journal{\PRL}{102}{096406}{2009}.


\bibitem{Kitaoka1998}
Y. Kitaoka, T. Kobayashi, A. Koda, H. Wakabayashi, Y. Niino, H. Yamakage, S. Taguchi, K. Amaya, K. Yamaura, M. Takano, A. Hirano, and R. Kanno,
\journal{\JPSJ}{67}{3703}{1998}.


\bibitem{Krimmel2005}
A. Krimmel, M. M\"ucksch, V. Tsurkan, M. M. Koza, H. Mutka, and A. Loidl,
\journal{\PRL}{94}{237402}{2005}.




\bibitem{Rokhsar1988}
D.~S.~Rokhsar and S.~A.~Kivelson,
\journal{\PRL}{61}{2376}{1988}.

\bibitem{Moesner2001}
R.~Moessner and S.~L.~Sondhi,
\journal{\PRL}{86}{1881}{2001}.


\bibitem{Nasu2013}
J. Nasu and S. Ishihara,
\journal{\PRB}{88}{094408}{2013}.





\bibitem{Zhou11}
H.~D.~Zhou, E.~S.~Choi, G.~Li, L.~Balicas, C.~R.~Wiebe, Y.~Qiu, J.~R.~D.~Copley, and J.~S.~Gardner,
\journal{\PRL}{106}{147204}{2011}.

\bibitem{Nakatsuji12}
S.~Nakatsuji, K.~Kuga, K.~Kimura, R.~Satake, N.~Katayama, E.~Nishibori, H.~Sawa, R.~Ishii, M.~Hagiwara, F.~Bridges, T.~U.~Ito, W.~Higemoto, Y.~Karaki, M.~Halim, A.~A.~Nugroho, J.~A.~Rodriguez-Rivera, M.~A.~Green, and C.~Broholm,
\journal{Science}{336}{559}{2012}.


\bibitem{Quilliam12}
J.~A.~Quilliam, F.~Bert, E.~Kermarrec, C.~Payen, C.~Guillot-Deudon, P.~Bonville, C.~Baines, H.~Luetkens, and P.~Mendels,
\journal{\PRL}{109}{117203}{2012}.


\bibitem{Ishiguro2013}
Y.~Ishiguro, K.~Kimura, S.~Nakatsuji, S.~Tsutsui, A.~Q.~R.~Baron, T.~Kimura, and Y.~Wakabayashi,
\journal{Nat.~Commum.}{4}{2022}{2013}.


\bibitem{Katayama2014}
N. Katayama, K. Kimura, Y. Han, J. Nasu, N. Drichko, Y. Nakanishi, M. Halim, Y. Ishiguro, R. Satake, E. Nishibori, M. Yoshizawa, T. Nakano, Y. Nozue, Y. Wakabayashi, S. Ishihara, M. Hagiwara, H. Sawa, and S. Nakatsuji,
arXiv:1403.4779. 




\bibitem{Corboz}
P. Corboz, M. Lajk\'o, A. M. L\"auchli, K. Penc, and F. Mila,
\journal{Phys. Rev. X}{2}{041013}{2012}.

\bibitem{Smerald2014}
A. Smerald and F. Mila,
\journal{\PRB}{90}{094422}{2014}.




\bibitem{Bersuker_text}
I. B. Bersuker,
{\it The Jahn-Teller Effect}
(Cambridge University Press, Cambridge, 2006).


\bibitem{Koizumi1999}
H. Koizumi and I. B. Bersuker,
\journal{\PRL}{83}{3009}{1999}.



\bibitem{Ralko2009}
A. Ralko, M. Mambrini, and D. Poilblanc,
\journal{\PRB}{80}{184427}{2009}.


\bibitem{Moesner2001a}
R. Moessner, S. L. Sondhi, and P. Chandra,
\journal{\PRB}{64}{144416}{2001}.



\bibitem{Djordjevic1982}
 Z. V Djordjevic, H. E. Stanley, and A. Margolina,
\journal{J. Phys. A. Math. Gen.}{15}{L405}{1982}.







\end{thebibliography}


\end{document}